\documentclass[prl, twocolumn, floatfix]{revtex4}
\setcitestyle{super}

\usepackage{graphicx}
\usepackage{amsmath}
\usepackage{color}

\begin{document}

\title{Phase Diagram of spatiotemporal instabilities in a large magneto-optical trap}

\author{M. Gaudesius$^{1}$, Y.-C. Zhang$^{2,3}$, T. Pohl$^{2}$, R.\ Kaiser$^{1}$, and G.\ Labeyrie$^{1}$\footnote{To whom correspondence should be addressed.}}
\affiliation{$^{1}$Universit\'{e} C\^{o}te d'Azur, CNRS, Institut de Physique de Nice, 06560 Valbonne, France}
\affiliation{$^{2}$Center for Complex Quantum Systems, Department of Physics and Astronomy, Aarhus University, DK-8000 Aarhus C, Denmark}
\affiliation{$^{3}$Shaanxi Key Laboratory of Quantum Information and Quantum Optoelectronic Devices, School of Physics, Xi'an Jiaotong University, Xi'an 710049, People's Republic of China}

\begin{abstract}

Large clouds of cold atoms prepared in a magneto-optical trap are known to present spatiotemporal instabilities when the frequency of the trapping lasers is brought close to the atomic resonance. This system bears similarities with trapped plasmas where the role of the Coulomb interaction is played by the exchange of scattered photons, and astrophysical objects such as stars whose size is dependent on radiative forces. We present in this paper a study of the phase-space of such instabilities, and reveal different dynamical regimes. Three dimensional simulations of the highly nonlinear atomic dynamics permit a detailed analysis of the experimental observations.

\end{abstract}

\maketitle

\section{I. Introduction}

Since its first realization in 1987~\cite{raab1987}, the magneto-optical trap (MOT) has been a very successful device for the preparation of dilute and cold atomic ensembles, and is nowadays present in many research labs throughout the world. In addition, in the so-called ''multiple scattering regime'' of the MOT~\cite{walker1990}, the nonlinear dynamics arising from the light-induced atomic interactions has attracted broad fundamental interest~\cite{labeyrie2006, pohl2006, mendonca2008, mendonca2012} and bears intriguing analogies to one-component plasmas~\cite{hansen1975}. Compared to other related systems such as Cepheid stars~\cite{cox1980}, the MOT presents the advantage to be highly tunable: the number of trapped particles can be adjusted, as well as the parameters that determine the magnitude of the effective charge.

spatiotemporal instabilities in various MOT geometries have been studied for two decades~\cite{wilkowski2000, diStephano2004, labeyrie2006}. We investigate in this work the instabilities of the ''balanced'' MOT, where the three pairs of laser beams used to generate the trapping and cooling force in three spatial dimensions are produced with six \textit{independent}, intensity-balanced laser beams (as opposed to three retro-reflected beams). In a previous work~\cite{gaudesius2020}, we specifically addressed the issue of the instability threshold which was experimentally studied both at fixed trapped atom number $N$ and as a function of $N$. The observed behaviors were compared to the predictions of different models. In the present paper, we are interested in analyzing the cloud's deformation modes in the unstable regime. We construct a phase-space diagram of these modes by scanning two of the main MOT parameters, the laser detuning and the magnetic field gradient. We identify several distinct regimes for the unstable MOT, with a particularly interesting regime occurring at low values of the magnetic field gradient that generates the MOT's confinement. The observed behaviors are compared to three-dimensional numerical simulations. 

\section{II. Experimental Setup and Numerical Simulations}

The characteristics of our $^{87}$Rb MOT have been described elsewhere~\cite{camara2014}. Its main feature is the ability to trap large atom numbers (up to 10$^{11}$), which enables us to reach a regime where collective effects are very strong. Producing such large cold atom samples is achieved by combining a MOT loading from an ambient room-temperature vapor with a large trapping volume. This volume is defined by the intersection of six independent laser beams of waist 3.4 cm that are intensity balanced. The typical peak intensity is 5 mW/cm$^2$ per beam, with a total power of $\approx$ 550 mW in the MOT. A pair of anti-Helmoltz coils produces the magnetic field gradient needed to generate the restoring force of the trap, whose typical value is 7 G/cm along the axis of the coils. Three pairs of compensation coils allow to finely tune the position of the zero of the magnetic trap, defining the MOT center. The typical loading time constant of the MOT is 2 s.

\begin{figure*}
\begin{center}
\includegraphics[width=2.0\columnwidth]{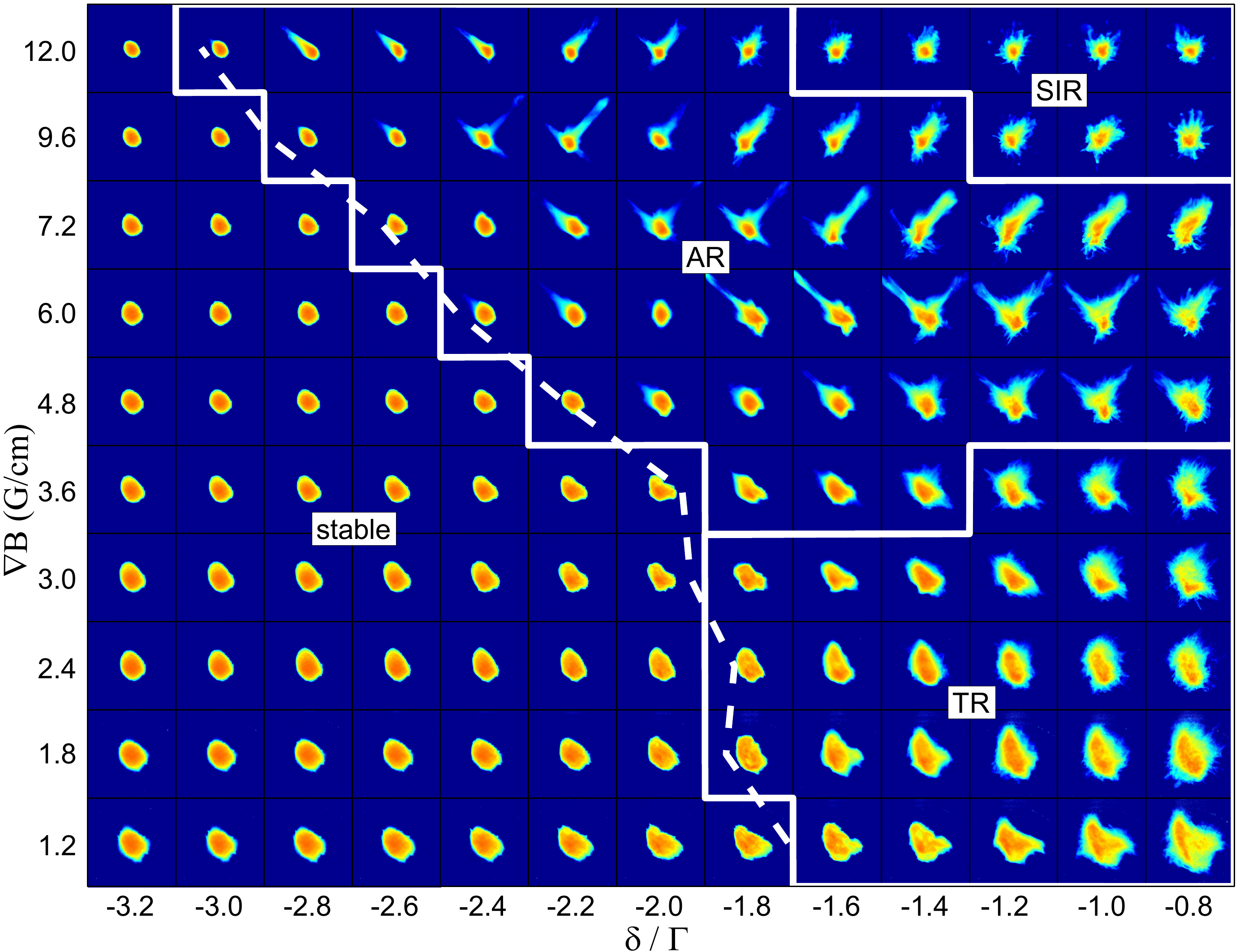}
\caption{($\delta$, $\nabla$B) instability phase diagram (experiment). The atom number is fixed at N = 1.5 $\times 10^{10}$. Each image corresponds to the average of fifty single-shot fluorescence images, with intensity displayed in logarithmic scale (the intensity range is 2 decades). The field of view is $6.5$ cm $\times~6.5$ cm. The dashed line materializes the instability threshold. The three different instability regimes (see text) are roughly delineated by the bold boxes.}
\label{map_avg}
\end{center}
\end{figure*}

Our aim is to study the instability phase diagram when two of the MOT parameters, the laser detuning $\delta$ and the magnetic field gradient $\nabla$B, are varied while the others (the laser intensity $I$ and trapped atom number $N$) are kept constant. Since the steady-state population in the trap depends on $\delta$ and $\nabla$B (see e.g.~\cite{camara2014}), we have to use the temporal sequence described in Ref.~\cite{gaudesius2020}. It consists of two successive phases, a ''loading'' phase were the parameters (here $\delta$ and the loading time) are adjusted to obtain the desired $N$, and an ''instability'' phase where $\delta$ is rapidly changed to a target value. The magnetic field gradient is set to the desired value and kept constant during the whole sequence. A fluorescence image is recorded at the end of this phase, using the MOT laser beams with a fixed detuning of -8 $\Gamma$ (where $\Gamma$ is the natural width of the employed $F = 2 \rightarrow F^{\prime} = 3$ transition) to minimize multiple scattering. The duration of the instability phase is 100 ms, long enough to decorrelate the cloud dynamics from the initial kick and short enough so that the MOT loading during the instability phase can be neglected. We probe the dynamics of the cloud in the instability phase by repeating this temporal sequence many times.

Several analytical models have been proposed in the past to predict the threshold of spatial instabilities in MOTs~\cite{labeyrie2006, tercas2010, mendonca2012}. They all rely on different assumptions using a macroscopic Ansatz and do not predict the same thresholds. In order to be less dependent on such simplifying assumptions, we have developed a numerical simulation, which in principle should include the various analytical predictions. In contrast to a first numerical approach described in Ref.~\cite{pohl2006} which confirmed one of the analytical models, we here use a three dimensional simulation based on microscopic ingredients of atom-light interactions~\cite{gaudesius2020}. It includes the usual Doppler trapping and cooling force, plus the two collective forces that are known to play an important role in the multiple scattering regime of the MOT~\cite{walker1990}. The first one is the compressive force due to the attenuation of the MOT's laser beams through the cloud (the so-called ''shadow effect'')~\cite{dalibard1988, walker1990}. The second one is the repulsive, effective Coulombian interaction force arising from exchange of scattered photons between trapped atoms~\cite{walker1990}. It is the interplay between these two collective forces of opposite sign that gives rise to the unstable behavior of the trapped atomic cloud. Some details about the model and simulation are provided in Section III.b of Ref.~\cite{gaudesius2020}. In this work, the reported numerical results were in qualitative agreement with experimentally-observed scaling laws, which gave us some confidence that the relevant physical ingredients were correctly included.

\section{III. Experimental Observations}

\begin{figure}
\begin{center}
\includegraphics[width=1\columnwidth]{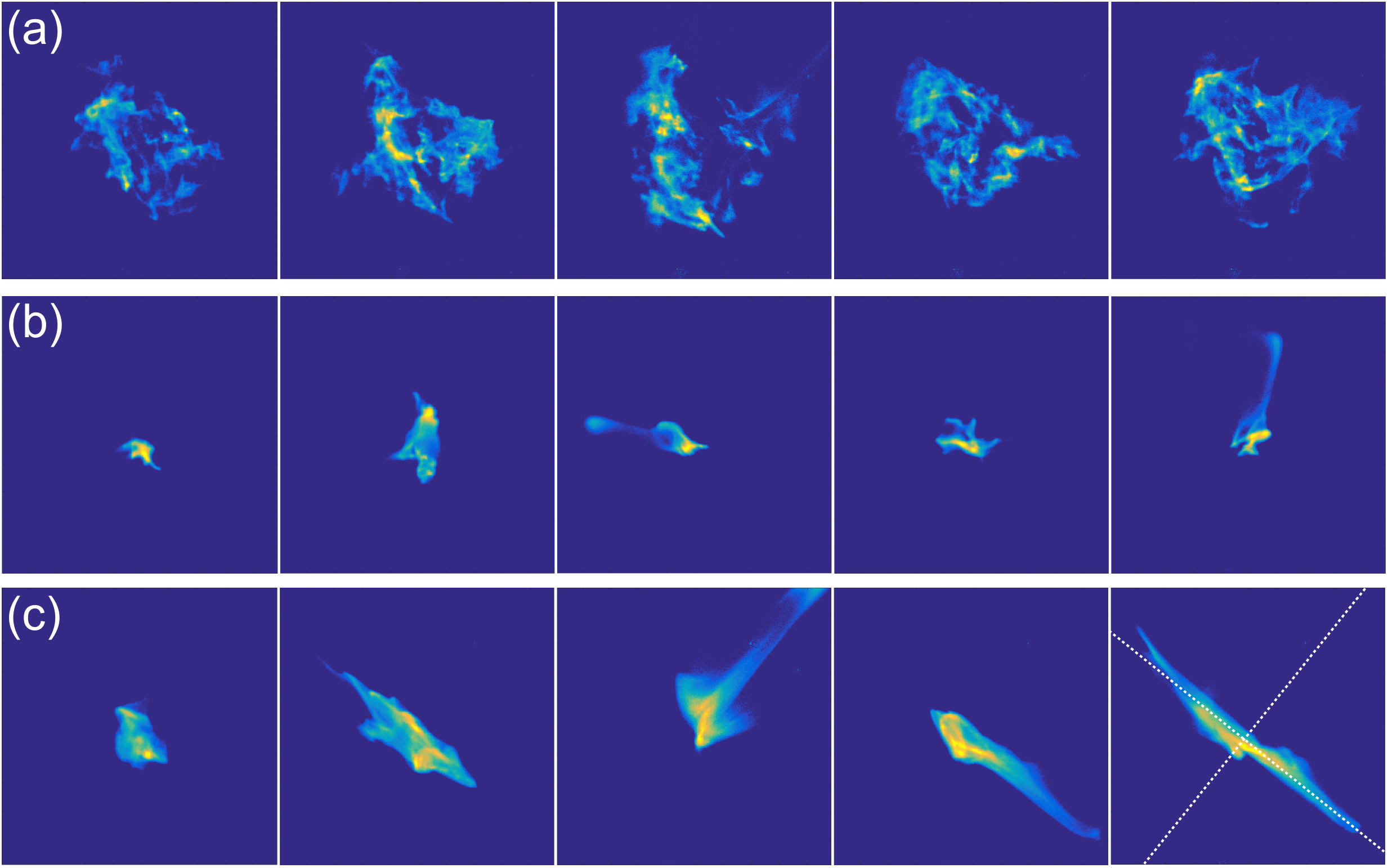}
\caption{Different instability regimes (experiment). We show single-shot images illustrating the cloud's dynamics in the three regimes identified in Fig.~\ref{map_avg}. (a): ''turbulent regime'' ($\nabla B = 1.2$ G/cm, $\delta = -\Gamma$). (b): ''statistically isotropic regime'' ($\nabla B = 12$ G/cm, $\delta = -\Gamma$). (c): ''anisotropic regime'' ($\nabla B = 7.2$ G/cm and $\delta = -1.8 \Gamma$). The dotted lines in the bottom right image show the directions of two pairs of MOT beams.}
\label{3_regimes}
\end{center}
\end{figure}

We report in Fig.~\ref{map_avg} the $\delta$ - $\nabla$B phase diagram obtained with our setup. The images shown are fluorescence images acquired using the procedure described in Section II, averaged over 50 cycles. The fluorescence is displayed in logarithmic scale (two orders of magnitude of dynamics). The axis of the magnetic field gradient coils (with one pair of trapping beams along it) is orthogonal to the plane of the figure. In this plane, the directions of the two other pairs of trapping beams correspond roughly to the diagonals of the images (see Fig.~\ref{3_regimes}).

The dashed line materializes the position of the instability threshold~\cite{gaudesius2020}. The cloud is stable in the region to the left of this line, and unstable to the right. We observe three broad instability regimes, approximately delineated by the bold boxes. Even though one can clearly distinguish qualitatively different instabilities, the transition between these regimes is not abrupt and a precise boundary is thus delicate to locate. We stress that although the details of the dynamics in the different regimes may vary with experimental conditions (beam alignment, ...), the general features and the separation of phase-space in three regions (four with the stable regime) is quite robust and reproducible.

The most striking new regime is observed in the weak $\nabla$B limit. Here the envelope of the cloud remains roughly stationary, but with small-scale structures inside undergoing a complex dynamics. As this regime appears to exhibit several small scale structures, we call it the ''turbulent regime'' (TR), noting that a comprehensive study in terms of weak turbulence would require a systematic study of scaling laws, going beyond the scope of this paper.

Another regime occurs at large $\nabla$B ($\geq$ 8 G/cm) and close to resonance. There the cloud shows large, anisotropic deformations, but in random directions. We thus call this the ''statistically isotropic regime'' (SIR). Everywhere else in the unstable region of the diagram, the cloud experiences deformations with a significant proportion along the MOT beams directions. This is the ''anisotropic regime'' (AR). Some examples of single-shot images illustrating the cloud's dynamics in these three regimes and their features described above are provided in Fig.~\ref{3_regimes}.

\begin{figure*}
\begin{center}
\includegraphics[width=2.0\columnwidth]{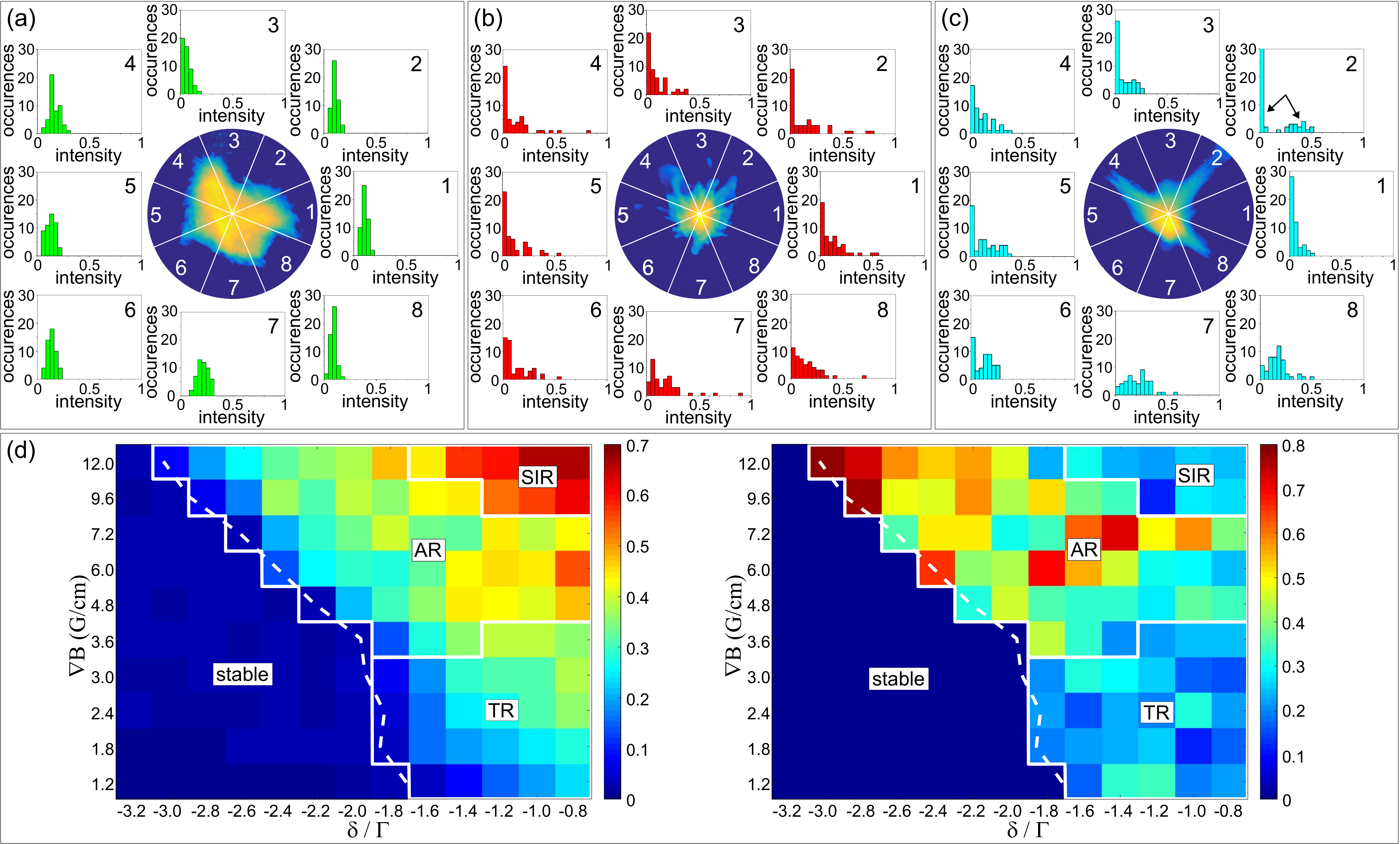}
\caption{Characterization of the three instability regimes using the eight angular sectors method (experiment). We use the same examples for the three regimes of instability as in Fig.~\ref{3_regimes}. (a): TR ($\nabla B = 1.2$ G/cm, $\delta = -\Gamma$), (b): SIR ($\nabla B = 12$ G/cm, $\delta = -\Gamma$) and (c): AR ($\nabla B = 7.2$ G/cm and $\delta = -1.8 \Gamma$). We show for each set of parameters the average image (intensity in logarithmic scale) and the histograms of the intensities in each sectors. In (d), we show color maps of the mean histogram width (left) and histogram width dispersion (right). The bold lines delineate the instability threshold (dashed line) and various regimes (solid lines) as in Fig.1.}
\label{hists}
\end{center}
\end{figure*}      

Although these three regimes of cloud deformation are qualitatively quite different, it seems useful to provide a more quantitative characterization. A well-known technique is Principal Component Analysis (PCA)~\cite{jolliffe2002}, which gives access to a set of uncorrelated spatial patterns involved in the dynamics. In our case, however, the number of such patterns is usually large and the information is too rich to characterize the different regimes in a simple way. We thus employ a cruder method, based on the division of each of the fifty images in a given data set into eight angular sectors as illustrated in Fig.~\ref{hists}. Note that one could in principle choose a smaller number of sectors, but that would result in a resolution too poor for our purpose. The center of each image is chosen as the center-of-mass of the average of the fifty realizations. We then compute the histograms of the intensities contained in each sector. The horizontal axis in the histograms shown in Fig.~\ref{hists} is the intensity in the corresponding sector, normalized to the total intensity in the image (sum of the eight sectors). The vertical axis is the number of occurrences for a given intensity. The total number of occurrences is equal to the number of images (fifty). This method of analysis provides some information on the magnitude of the cloud's deformation and its anisotropy. Note, however, that this approach does not detect shape-preserving oscillations, whose histograms would not be distinguishable from a that of stable clouds.

The data shown in Fig.~\ref{hists} correspond to the same sets of parameters as in Fig.~\ref{3_regimes}. In (a) (TR), the histograms are similar in all sectors, bell-shaped with approximately the same mean and width, which is rather small (at most $30\%$ of the total in one sector). These observations are consistent with limited deformations of the cloud's envelope, equivalent in all directions. The complex dynamics of the small-scale structures seen in Fig.~\ref{3_regimes}(a) is averaged by the relatively large area of the sectors. In (b) (SIR), we observe that the histograms are also more or less similar in all sectors. However, their shape and width are very different from case (a). It is seen that the histograms correspond to a decreasing probability law, with the most probable intensity in each sector being almost zero. This means that the cloud undergoes large deformations associated with large excursions of its center of mass. Concurrently, the large width of the histograms, with instances where $90\%$ of the cloud is in one sector, indicates very anisotropic deformations. In case (c) (AR), the situation is again quite different. Here, the shapes and widths are different in all sectors. In particular, we observe double-peaked histograms such as in sector 2 (arrows). This feature indicates that the cloud jumps repeatedly between two preferred states with different shapes, one of which is highly elongated along the direction of one (or several) MOT beam. Overall, the cloud deformations are quite large with up to $50\%$ of the cloud in one sector. We thus see that based on the characteristics of the collection of eight histograms, namely the dispersion of histograms widths and value of the average width, we can establish a distinction between the different instability regimes. This is confirmed by the color maps shown in (d), with the mean width displayed on the left and the dispersion of the widths on the right. The bold boxes delineating the different regimes, obtained by visual inspection of the sets of fifty images, are the same as in Fig.~\ref{map_avg}. As can be seen, the result of the eight sectors analysis is in fair agreement with the visual inspection: in the TR both mean and dispersion are small, for the IR the mean is large and the dispersion small, and in the AR the mean is intermediate but most importantly the dispersion is large.

\section{IV. Numerical results}

We now investigate the deformation regimes observed with the 3D simulation scheme, aiming at a direct comparison between our microscopic model and the experiment. As will be seen, the modeling in three spatial dimensions appears essential as it allows for deformation regimes that were not accessible in previous studies (see e.g.~\cite{pohl2006}).

The numerical simulation proceeds as follow. We start at t = 0 with $7 \times 10^3$ ''super-particles'' (see Ref.~\cite{gaudesius2020}) with random positions within a Gaussian density distribution. We then let the cloud evolve under the action of the forces described in Sec.II. In the unstable regime, we observe that the spontaneous cloud oscillations typically occur after a transient time $t_{tr}$ (see Fig.3 of Ref.~\cite{gaudesius2020}) which varies with the parameters, here $\delta$ and $\nabla$B. We then randomly choose 50 time values $> t_{tr}$, and integrate the corresponding 3D density distributions along the z axis to allow for a direct comparison with experimental images. We adopt this approach, where we probe different times for a single realization (while in the experiment we probe different realizations), because the numerical simulations are too time-consuming to sample many realizations. By comparing the results of these different protocols, we implicitly assume ergodicity. 

\begin{figure}
\begin{center}
\includegraphics[width=1\columnwidth]{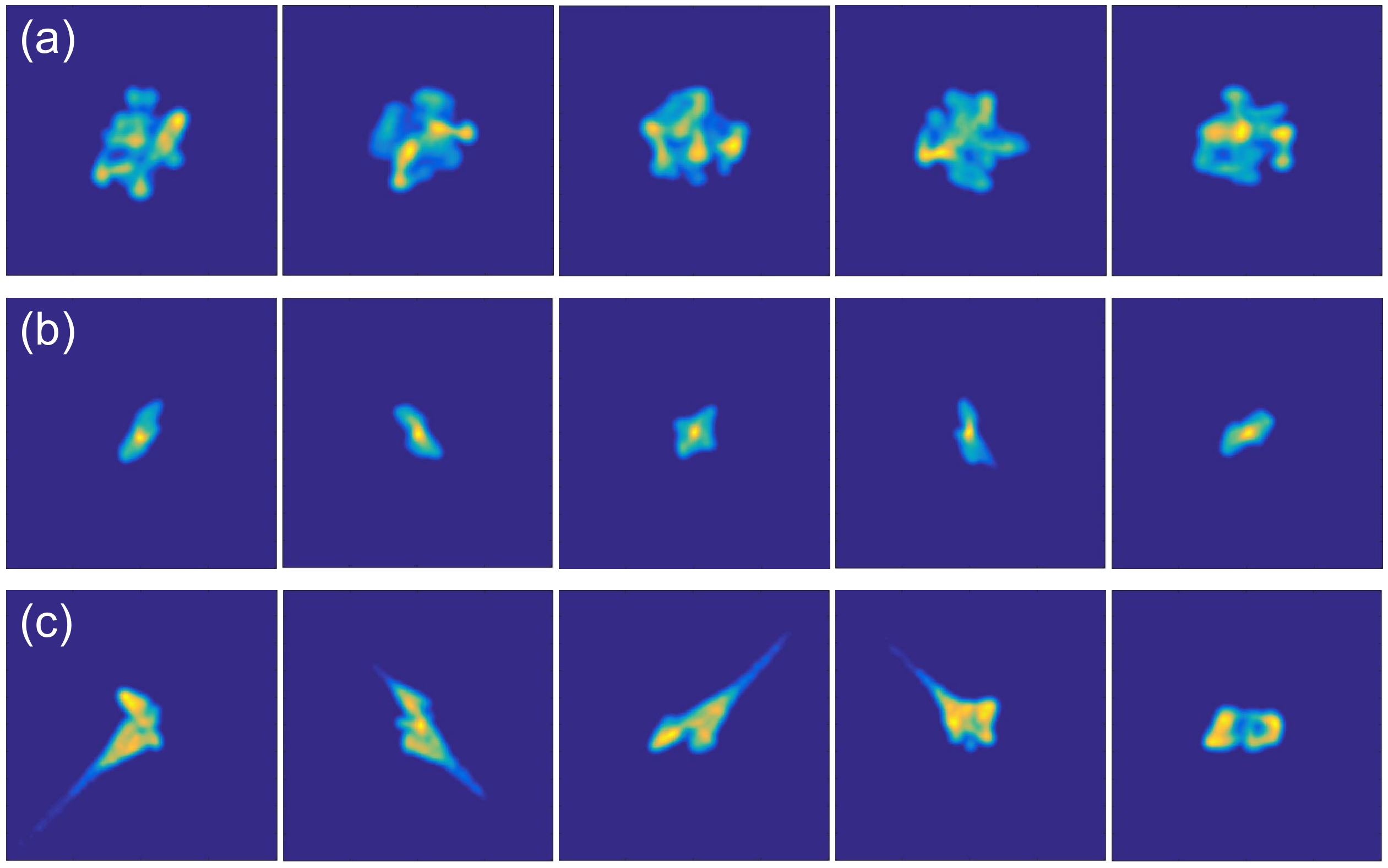}
\caption{Different instability regimes (simulations). We illustrate three of the instability regimes obtained in the simulations that are similar to those observed in the experiment (see Fig.~\ref{3_regimes}). The images in each line correspond to different times in the simulation. (a): $\nabla$B = 1.2 G/cm, $\delta = -1.3 \Gamma$. (b): $\nabla$B = 12 G/cm, $\delta = -2.6 \Gamma$. (c): $\nabla$B = 4.8 G/cm, $\delta = -1.9 \Gamma$.}
\label{sim_regimes}
\end{center}
\end{figure}

We report in Fig.~\ref{sim_regimes} some examples of instability regimes obtained in the simulations that are reminiscent of those observed in the experiment. In particular, the typical density distributions as well as the histograms obtained with the eight sectors method match fairly well the experimental ones. The first regime (a), also occurring for low magnetic field gradients, is characterized by small-sized structures fluctuating inside a roughly fixed envelope. The structures are not as small and numerous as in the experiment, which we attribute to the coarse-graining due to the limited number of super-particles in the simulations. In the second regime (b), observed in a limited range of detunings and for high B-field gradients, the cloud undergoes large deformations in arbitrary directions similarly to the SIR of Fig.~\ref{3_regimes}. In the third regime (c), the cloud's deformations are mostly occurring along the directions of the MOT's laser beams (the diagonals of the images) with the cloud jumping from side to side in a manner very similar to what is experimentally observed in the AR (Figs.2(c), 3(c) and 4(c)). This is an important observation, which shows that this kind of spontaneous symmetry breaking leading to large c.o.m excursions can occur in a perfectly centro-symmetric trapping potential, in the absence of experimental imperfections such as local intensity imbalance or beam misalignment.

\begin{figure}
\begin{center}
\includegraphics[width=1\columnwidth]{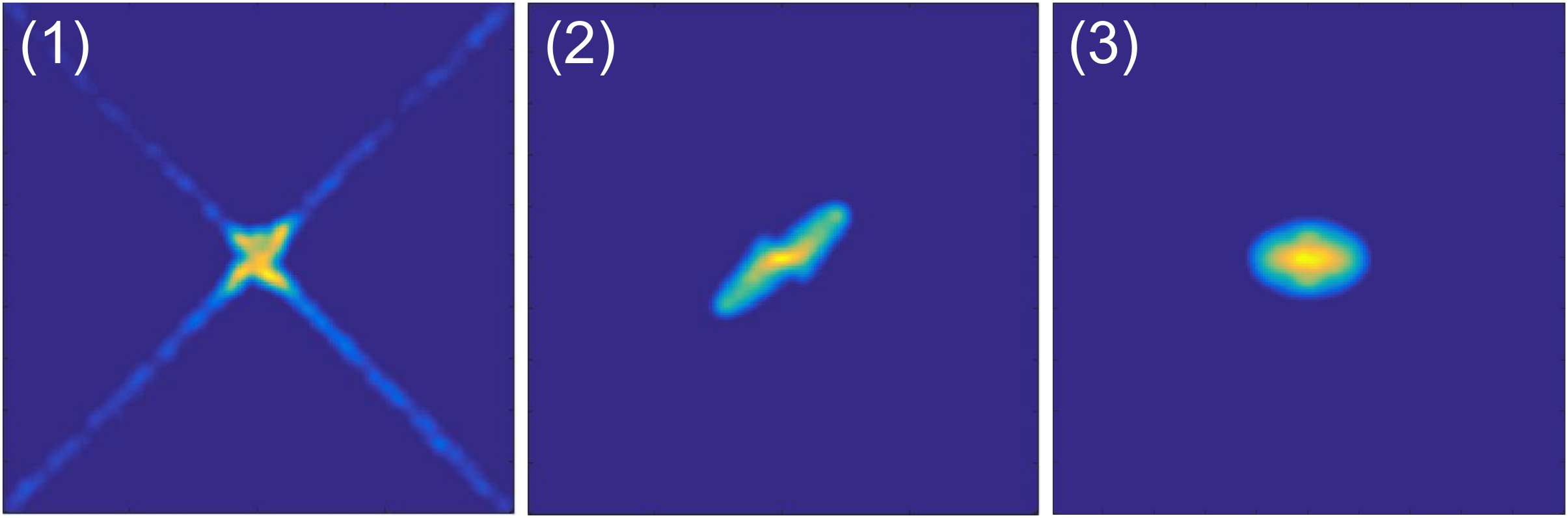}
\caption{Other instability regimes (simulations). These instability regimes are not observed in the experiment. (1): $\nabla$B = 7.2 G/cm, $\delta = -1.1 \Gamma$. (2): $\nabla$B = 9.6 G/cm, $\delta = -2.9 \Gamma$. (3): $\nabla$B = 6 G/cm, $\delta = -3.3 \Gamma$.}
\label{other_regimes}
\end{center}
\end{figure}

Thus, in addition to being able to reproduce the scaling laws observed for the instability threshold~\cite{gaudesius2020}, our model and numerical simulations do predict some instability regimes that are similar to what is observed experimentally. We note, however, that the regions of phase-space where these regimes occur are not in perfect match with the experiment. The simulations also predict modes of instabilities that could not be observed in the current experiment (see Fig.~\ref{other_regimes}). The regime shown in (1) occurs for all values of $\nabla$B when the laser frequency is brought close enough to resonance. Typically, the cloud in this regime exhibits two components: a central core which shows spatial and temporal fluctuations, and a nearly static ''cross'' of atoms accumulating on the axes of the laser beams. There is a continuous transition from the anisotropic regime of Fig.~\ref{sim_regimes}(c) to this regime when $\left| \delta \right|$ is decreased. The regime shown in Fig.~\ref{other_regimes}(2) occurs at relatively large $\nabla$B and $\left|\delta\right|$. It is characterized by a ''twisted'' elliptical shape oriented along the axis of one pair of MOT beams. Different simulation runs can yield different orientations. In this regime, the spatial fluctuations of the cloud are reduced. Finally, the regime in (3) occurs at intermediate to large $\nabla$B, for detunings just above threshold. There, the cloud assumes an almost elliptical shape, horizontal or vertical. In this regime, the cloud is nearly static but the trapping symmetry is spontaneously broken. The precise origin of this different behavior in the simulations and experiments remains to be clarified in future numerical work to explore a broader parameter space and scrutinize some of the remaining approximations employed in the simulations.     

\section{Conclusion}
We presented in this paper an experimental study of the spatiotemporal instabilities taking place in a balanced magneto-optical trap containing a large number of atoms. By scanning the laser detuning and magnetic field gradient at fixed atom number, we identified three broad instability regimes. These observations thus constitute a finer description of unstable MOTs using a characterization based on spatial symmetries. We have identified new regimes, beyond previous predictions. 
These we also observed in extensive three-dimensional numerical simulations. In the most spectacular instability regime observed in the experiments at low $\nabla$B, the cloud breaks into small-scale filaments undergoing complex motions reminiscent of turbulence e.g. in flames. We plan to investigate this regime in more details, to determine if turbulence is indeed present in this system and of what nature. Making use of the high tunability of the MOT may allow us to induce transitions between different types of turbulence. This ab initio study of optomechanical instabilities contributes to the wider field of collective effects in light-matter coupling and we speculate that some of the features identified in our experiments and simulations can be present in plasmas and astrophysical systems.  

\section{Acknowledgements}
This work was performed in the framework of the European Training Network ColOpt, which is funded by the European Union (EU) Horizon 2020 program under the Marie Sklodowska-Curie action, grant agreement No. 721465, and in the framework of the European project ANDLICA, ERC Advanced grant No. 832219. It has also been supported by the Danish National Research Foundation through a Niels Bohr Professorship to TP and through the Center of Excellence ''CCQ'' (grant agreement No. DNRF156).

\end{document}